\documentclass{article}
\usepackage{interspeech2008,amssymb,amsmath,epsfig}
\ninept
\def\reg{{\rm\ooalign{\hfil
     \raise.07ex\hbox{\scriptsize R}\hfil\crcr\mathhexbox20D}}}

\newcommand{\A}{\textsc{aurora-2}~}

\title{TR02: State dependent oracle masks for improved dynamical features}

\makeatletter
\def\name#1{\gdef\@name{#1\\}}
\makeatother
\name{{\em J. F. Gemmeke and B. Cranen}}

\address{Dept. of Linguistics, Radboud University, Nijmegen, The Netherlands \\
{\small \tt \{J.Gemmeke, B.Cranen\}@let.ru.nl}}

\begin{document}
\maketitle
\begin{abstract}
Using the AURORA-2 digit recognition task, we show that recognition accuracies obtained with classical, SNR based oracle masks can be substantially improved by using a state-dependent mask estimation technique.
\end{abstract}

\section{Introduction}\label{sec:intro}
In Missing Data Techniques (MDT) for noise robust automatic speech recognition (ASR), it is often implicitly assumed that using an SNR based oracle mask\,\footnote{These oracle masks are computed by comparing spectro-temporal representations of the underlying speech and noise signals. Features dominated by speech energy are dubbed reliable; features dominated by noise energy unreliable.} guarantees maximum recognition accuracy. Generally speaking, however, this is not necessarily true.
\par
In previous work \cite{Gemmeke2008}, the authors showed that the portions of an oracle mask which are important for recognition accuracy are speech sound
dependent. In this paper we exploit this finding by a state dependent treatment of reliable features. Using a different mask estimator for every state in an HMM model and selecting the mask estimator for each time frame based on an externally provided state transcription, we generate a new type of oracle mask. In this paper we compare recognition accuracies obtained with state-dependent oracle masks and classical oracle masks on the \A digit recognition task.

\section{Experiments and results}\label{sec:experiments}
Experiments on test set A of the \A digit corpus were carried out using a MATLAB HMM-based MDT recognizer in which the masks for delta coefficients were computed as the delta's of the static masks (cf. \cite{Van2006} for implementation and model details). Our state-dependent mask estimator was based on binary SVM classifiers using LIBSVM.
\par
We trained separate SVM-models for all $S=179$ HMM  states and all $K=23$ Mel frequency bands, resulting in $S \times K = 4117$ models. The frame-based SVM features we used consisted of 'Subband Energy to Subband Noise Floor Ratio' and 'Flatness' as in \cite{Seltzer2004}, the harmonic and random components of the noisy speech signal \cite{Van2004a} and the noisy speech acoustic vectors. Reliability labels used in training were obtained from the (classical) oracle mask. Every state-specific SVM mask estimator was trained on the frames from the multi-condition train set, which were assigned to the same state by a forced alignment of the corpus utterances with the reference transcription. All $4117$ models were trained with the same SVM-feature vector.
\par
Table \ref{table:digitrecog} shows the recognition accuracies for the classical and the state-dependent oracle mask, as well as the accuracy gain.
\par
\begin{center}
\begin{table}
\vspace*{-2.5ex}
\caption{\label{table:digitrecog}\A digit recognition accuracy (\%).}
\begin{tabular*}{8.0cm}{|l|p{4mm}p{4mm}p{4mm}p{4mm}p{4mm}p{4mm}p{4mm}|}
\hline
~ &\multicolumn{7}{|c|}{SNR}\\
method &\multicolumn{1}{|p{4mm}}{clean}
       &\multicolumn{1}{c}{20}
       &\multicolumn{1}{c}{15}
       &\multicolumn{1}{c}{10}
       &\multicolumn{1}{c}{ 5}
       &\multicolumn{1}{c}{ 0}
       &\multicolumn{1}{c|}{-5} \\
\hline
classical     &\multicolumn{1}{r}{\footnotesize 99.7}
              &\multicolumn{1}{r}{\footnotesize 99.2}
              &\multicolumn{1}{r}{\footnotesize 99.3}
              &\multicolumn{1}{r}{\footnotesize 98.4}
              &\multicolumn{1}{r}{\footnotesize 96.3}
              &\multicolumn{1}{r}{\footnotesize 88.3}
              &\multicolumn{1}{r|}{\footnotesize 58.6} \\
state dep.    &\multicolumn{1}{r}{\footnotesize 99.7}
              &\multicolumn{1}{r}{\footnotesize 99.5}
              &\multicolumn{1}{r}{\footnotesize 99.5}
              &\multicolumn{1}{r}{\footnotesize 99.2}
              &\multicolumn{1}{r}{\footnotesize 97.9}
              &\multicolumn{1}{r}{\footnotesize 92.1}
              &\multicolumn{1}{r|}{\footnotesize 67.3} \\
$\Delta$ acc. &\multicolumn{1}{r}{\footnotesize  0.0}
              &\multicolumn{1}{r}{\footnotesize  0.3}
              &\multicolumn{1}{r}{\footnotesize  0.2}
              &\multicolumn{1}{r}{\footnotesize  0.8}
              &\multicolumn{1}{r}{\footnotesize  1.5}
              &\multicolumn{1}{r}{\footnotesize  3.8}
              &\multicolumn{1}{r|}{\footnotesize  8.7} \\

\hline
\end{tabular*}
\end{table}
\end{center}
\vspace*{-4ex}
\begin{minipage}[h]{9cm}
\parindent 5mm
\vspace*{-3ex}

\noindent Clearly, the state dependent method performs consistently better than the classical oracle mask, with larger gains in more adverse conditions.

\section{Discussion and Conclusions}\label{sec:conclusions}
The classical, SNR based oracle mask only describes which static coefficients are reliable. Since treatment of dynamic features is missing data decoder specific the classical oracle mask is not necessarily the 'ideal' mask.
Detailed analysis revealed that state dependent masks contain fewer isolated reliable elements than classical ones. In our setup (but also in those of others, e.g., \cite{Demange2006}) coarser granularity is beneficial for recognition performance, because isolated reliable mask coefficients can result in delta mask coefficients that are mistakenly labeled reliable.
\par
Our findings might also be useful for speech decoding without a priori information about the state sequence. Oracle recognition accuracy would theoretically come within reach if, for each frame, one can afford to evaluate as many mask vectors as there are states (i.e. 179 in case of \A).

This is a significant reduction of complexity as compared to the $2^{23}=8.388.608$ theoretically possible masks and without the loss of accuracy reported in \cite{Demange2006}. For small vocabulary tasks, today's computing power might even make a brute force approach feasible.
\par
Our future research, however, will focus on a further reduction of computational complexity by exploiting state transition constraints.

\section{Acknowledgements}
The research of Jort Gemmeke was carried out in the {MIDAS} project, granted under the Dutch-Flemish STEVIN program.

\eightpt
\bibliographystyle{IEEEtran}
\bibliography{Interspeech2008}\label{sec:ref}
\end{minipage}

\end{document}